\documentclass[preprint,showpacs,showkeys,preprintnumbers,amsmath,amssymb]{revtex4}
 \usepackage{dcolumn}
 \usepackage{bm}

\begin{document}

 \title{ Fermions Tunneling from Apparent Horizon of FRW Universe }

 \author{Ran Li}

 \author{Ji-Rong Ren}

 \thanks{Corresponding author. Electronic mail: renjr@lzu.edu.cn}

 \author{Dun-Fu Shi}

 \affiliation{Institute of Theoretical Physics, Lanzhou University, Lanzhou, 730000, Gansu, China}

 \begin{abstract}
 In the paper [arXiv:0809.1554], the scalar particles' Hawking
 radiation from the apparent horizon of Friedmann-Robertson-Walker(FRW) universe
 was investigated by using the tunneling formalism.
 They obtained the Hawking temperature associated with the apparent
 horizon,
 which was extensively applied in
 investigating
 the relationship between the first law of thermodynamics and Friedmann
 equations.
 In this paper, we calculate Fermions' Hawking radiation
 from the apparent horizon of FRW
 universe via tunneling
 formalism. Applying WKB approximation to the
 general covariant Dirac
 equation in FRW spacetime background,
 the radiation spectrum and Hawking temperature of
 apparent horizon are correctly recovered, which supports
 the arguments presented in the paper [arXiv:0809.1554].

 \end{abstract}

 \pacs{04.62.+v, 04.70.Dy, 98.80.Jk.}

 \keywords{Fermions, tunneling, Hawking radiation, FRW Universe.}

 \maketitle

 In the recent paper \cite{cai1}, Cai \textit{et al} discussed the
 scalar particles' Hawking radiation from the apparent horizon of
 FRW universe, which revealed that there is a Hawking temperature
 associated with the apparent horizon. In this paper, we will
 investigate Fermions' tunneling from the apparent horizon of FRW
 universe. The results in the present paper together with that in \cite{cai1}
 fill in the gap existing in the literatures investigating the
 relationship between the first law of thermodynamics and Friedmann
 equations of FRW universe.

 Ted Jacobson\cite{jacobson} has been able to derive Einstein equation from
 the proportionality of entropy and horizon area
 together with the fundamental Clausius relation $\delta Q=TdS$
 connecting heat, entropy, and temperature.
 The key idea is to demand that this relation holds for all the local
 Rindler causal horizons through each spacetime point, with $\delta Q$
 and $T$ interpreted as the energy flux and Unruh
 temperature seen by an accelerated observer just inside the horizon.
 This perspective suggests that Einstein field equation can be
 viewed as an equation of state of spacetime. The idea has also
 applied to $f(R)$ theory\cite{ted} and scalar-tensor
 theory\cite{cai2}, where the non-equilibrium thermodynamic must be
 taken into account.

 It has been proved that the above idea can also be applied in
 establishing the relationship between the Friedmann equations and
 the first law of thermodynamics in the framework of
 Friedmann-Robertson-Walker(FRW) universe.
 Assuming that the apparent horizon of FRW universe has the
 temperature $T=1/2\pi\tilde{r}_A$ and entropy $S=A/4$, where $\tilde{r}_A$
 and $A$ are the radius and the area of the apparent horizon
 respectively, the Friedmann equations can be derived from the
 Clausius relation\cite{cai3} and the first Friedmann equation can be cast into
 the form of the unified first law\cite{cai2,cai4}.
 For some related investigations see also\cite{akbar,akbar1,cai5,cai6,ge,gong,wu,zhu,akbar2}.

 However, whether there is a Hawking temperature associated with the
 apparent horizon of FRW universe is still an open question to investigate.
 In a recent paper \cite{cai1}, the scalar particles' Hawking
 radiation from the Apparent Horizon of FRW Universe
 was investigated by using the tunneling formalism.
 They obtained the Hawking temperature associated with the apparent
 horizon of FRW universe. They found that the Kodama observer inside
 the apparent horizon do see a thermal spectrum with temperature
 $T=1/2\pi\tilde{r}_A$, which is caused by particles tunneling from
 the outside apparent horizon to the inside apparent horizon.

 The semi-classical derivation of Hawking radiation as
 tunneling process\cite{parikh} was initially proposed by
 Parikh and Wilczek. In recent years, it has
 already attracted a lot of
 attention. In this method, the imaginary part of the action is
 calculated using the null geodesic equation.
 Zhang and Zhao extended this method to the charged
 Reissner-Nordstr\"{o}m black hole\cite{zhangjhep}
 and the rotating Kerr-Newman black hole\cite{zhangplb}.
 See also \cite{jiangwu} for a different discussion.
 M. Angheben \textit{et al} \cite{angheben} also proposed a
 derivation of Hawking radiation by calculating the particles'
 classical action from the Hamilton-Jacobi equation, which is an
 extension of the complex path analysis of T. Padmanabhan \textit{et
 al} \cite{padmanabhan}.
 Very recently, a new calculation concerning fermions'
 radiation from the stationary spherical symmetric black hole was
 done by R. Kerner and R. B. Mann in \cite{kernerarxiv}. This
 method has been generalized to the more general and complicated
 spacetime background\cite{liplb,lic,kerner,chen,chen1,jiang,jiang1} and
 dynamical black hole\cite{vanzo}.

 In this paper, we will calculate the Fermions' Hawking radiation
 from the apparent horizon of FRW
 universe via tunneling
 formalism. Applying WKB approximation to the general covariant Dirac
 equation in FRW spacetime background,
 we also obtain the Hawking temperature of
 apparent horizon of FRW Universe. Our results
 obtained by taking the fermion
 tunneling into account support
 the arguments presented in the previous paper \cite{cai1}.

 For convenience, we firstly give some results related to the FRW
 universe.
 The FRW metric is given by
 \begin{eqnarray}
 ds^2=-dt^2+a^2(t)\left(\frac{dr^2}{1-kr^2}+
 r^2(d\theta^2+\textrm{sin}^2\theta d\phi^2)\right)\;\;,
 \end{eqnarray}
 where $a(t)$ is the scale factor and $k=1$, $0$ and $-1$ represent
 the closed, flat and open universe respectively. Introducing
 $\tilde{r}=ar$, the FRW metric (1) can be rewritten as
 \begin{eqnarray}
 ds^2=h_{ab}dx^a dx^b+\tilde{r}^2d\Omega_2^2\;\;,
 \end{eqnarray}
 where $x^a=(t, r)$, $h_{ab}=\textrm{diag}(-1, \frac{a^2}{1-kr^2})$
 and $d\Omega_2^2$ represents the line element of $S^2$. The
 apparent horizon is defined by the equation
 \begin{eqnarray}
 h^{ab}\partial_a\tilde{r}\partial_b\tilde{r}=0\;\;,
 \end{eqnarray}
 which gives us the location of apparent horizon explicitly as
 \begin{eqnarray}
 \tilde{r}_A=\frac{1}{\sqrt{H^2+k/a^2}}\;\;,
 \end{eqnarray}
 with $H=\dot{a}/a$ being the Hubble parameter. For simplicity, we
 will use the $(t, \tilde{r})$ coordinates, in which the FRW metric can
 be rewritten as
 \begin{eqnarray}
 ds^2=-\frac{1-\tilde{r}^2/\tilde{r}_A^2}{1-k\tilde{r}^2/a^2}dt^2
 -\frac{2H\tilde{r}}{1-k\tilde{r}^2/a^2}dtd\tilde{r}
 +\frac{1}{1-k\tilde{r}^2/a^2}d\tilde{r}^2+\tilde{r}^2d\Omega_2^2\;\;.
 \end{eqnarray}
 The inverse of the metric of the $(t,\tilde{r})$ parts is given by
 \begin{eqnarray}
 g^{tt}=-1\;\;,\;\;g^{t\tilde{r}}=-H\tilde{r}\;\;,\;\;
 g^{\tilde{r}\tilde{r}}=1-\tilde{r}^2/\tilde{r}_A^2\;\;.
 \end{eqnarray}

 Now we calculate the Fermions' Hawking radiation from the apparent horizon
 of the FRW universe via the tunneling formalism. Let us start with
 the massless spinor field $\Psi$ obeyed the general covariant Dirac equation
 \begin{eqnarray}
 -i\hbar\gamma^a e_{a}^{\mu}\nabla_{\mu}\Psi=0\;\;,
 \end{eqnarray}
 where $\nabla_\mu$ is the spinor covariant derivative defined by
 $\nabla_\mu=\partial_\mu+\frac{1}{4}\omega_\mu^{ab}\gamma_{[a}\gamma_{b]}$,
 and $\omega_\mu^{ab}$ is the spin connection, which can be given
 in terms of the tetrad $e_a^\mu$.The $\gamma$ matrices are selected
 as
 \begin{eqnarray*}
 \gamma^0&=&\left(%
 \begin{array}{cc}
  i & 0 \\
  0 & -i \\
 \end{array}%
 \right)\;\;,\\
 \gamma^1&=&\left(%
 \begin{array}{cc}
  0 & \sigma^3 \\
  \sigma^3 & 0 \\
 \end{array}%
 \right)\;\;,\\
 \gamma^2&=&\left(%
 \begin{array}{cc}
  0 & \sigma^1 \\
  \sigma^1 & 0 \\
 \end{array}%
 \right)\;\;,\\
 \gamma^3&=&\left(%
 \begin{array}{cc}
  0 & \sigma^2 \\
  \sigma^2 & 0 \\
 \end{array}%
 \right)\;\;,
 \end{eqnarray*}
 where the matrices $\sigma^k(k=1,2,3)$ are the Pauli matrices.
 According to the line element (5) and the inverse metric given in
 (6), the tetrad field can be selected to be
 \begin{eqnarray*}
 e_0^\mu&=&(1, H\tilde{r}, 0, 0)\;,\\
 e_1^\mu&=&(0, \sqrt{1-k\tilde{r}^2/a^2}, 0, 0)\;,\\
 e_2^\mu&=&(0, 0, \tilde{r}^{-1}, 0)\;,\\
 e_3^\mu&=&(0, 0, 0, (\tilde{r}\textrm{sin}\theta)^{-1})\;.
 \end{eqnarray*}
 Without loss of generality, we employ the following ansatz for
 spinor field with the spin up
 \begin{eqnarray}
 \Psi=\left(%
 \begin{array}{c}
  A(t,r,\theta,\phi) \\
  0\\
  B(t,r,\theta,\phi) \\
  0
 \end{array}%
 \right)\textrm{exp}\big[\frac{i}{\hbar}I(t,r,\theta,\phi)\big]\;.
 \end{eqnarray}
 It should be noted that the spin-down case is just analogous.
 In order to apply the WKB approximation, we can insert the ansatz
 (8) for the spin-up spinor field $\Psi$ into the general covariant Dirac equation (7).
 Dividing by the exponential term and neglecting the terms with $\hbar$,
 one can arrive at the following four equations
 \begin{eqnarray}
 \left\{
  \begin{array}{ll}
  iA(\partial_t I+H\tilde{r}\partial_{\tilde{r}}I)+B\sqrt{1-k\tilde{r}^2/a^2}
  \partial_{\tilde{r}}I =0  \;,  \\
  B\left(\frac{1}{\tilde{r}}\partial_{\theta}I+
  \frac{i}{\tilde{r}\textrm{sin}\theta}\partial_\phi I\right)=0  \;,  \\
  A\sqrt{1-k\tilde{r}^2/a^2}\partial_{\tilde{r}}I-iB
  (\partial_t I+H\tilde{r}\partial_{\tilde{r}}I)=0  \;,  \\
  A\left(\frac{1}{\tilde{r}}\partial_{\theta}I+
  \frac{i}{\tilde{r}\textrm{sin}\theta}\partial_\phi I\right)=0 \;.
  \end{array}
 \right.
 \end{eqnarray}
 Note that although $A$ and $B$ are not constant, their
 derivatives and the spin connections are all of the factor
 $\hbar$, so can be neglected to the lowest order in WKB
 approximation.

 To carry out the separation of variables for the
 above equations, we must analysis the symmetries of the metric (5).
 For the metric (5), the Kodama vector is given by\cite{cai1}
 \begin{eqnarray}
 K^a=\sqrt{1-k\tilde{r}^2/a^2}\left(\frac{\partial}{\partial
 t}\right)^a\;\;.
 \end{eqnarray}
 The Kodama vector in dynamical spacetime is of the same significance with the Killing
 vector in static spacetime. It should be noted that the Kodama
 vector is time-like, null and space-like as
 $\tilde{r}<\tilde{r}_A$, $\tilde{r}=\tilde{r}_A$ and
 $\tilde{r}>\tilde{r}_A$ respectively. Using the Kodama vector,
 one can define the energy measured by the Kodama observer
 \begin{eqnarray}
 \omega=-K^a\partial_a I=-\sqrt{1-k\tilde{r}^2/a^2}\partial_t I\;\;.
 \end{eqnarray}
 Using the definition of energy, the classical action can be
 separated as
 \begin{eqnarray}
 I=-\int\frac{\omega}{\sqrt{1-k\tilde{r}^2/a^2}}dt+R(\tilde{r})+P(\theta,\phi)\;\;,
 \end{eqnarray}
 Substituting the above ansatz into Eq.(9) yields
 \begin{eqnarray}
 \left\{
  \begin{array}{ll}
  iA(-\frac{\omega}{\sqrt{1-k\tilde{r}^2/a^2}}+H\tilde{r}\partial_{\tilde{r}}I)+B\sqrt{1-k\tilde{r}^2/a^2}
  \partial_{\tilde{r}}I =0  \;,  \\
  B\left(\frac{1}{\tilde{r}}\partial_{\theta}P(\theta,\phi)+
  \frac{i}{\tilde{r}\textrm{sin}\theta}\partial_\phi P(\theta,\phi)\right)=0  \;,  \\
  A\sqrt{1-k\tilde{r}^2/a^2}\partial_{\tilde{r}}I-iB
  (-\frac{\omega}{\sqrt{1-k\tilde{r}^2/a^2}}+H\tilde{r}\partial_{\tilde{r}}I)=0  \;,  \\
  A\left(\frac{1}{\tilde{r}}\partial_{\theta}P(\theta,\phi)+
  \frac{i}{\tilde{r}\textrm{sin}\theta}\partial_\phi P(\theta,\phi)\right)=0 \;.
  \end{array}
 \right.
 \end{eqnarray}
 It is easy to see that $P$ must be a complex
 function, which means it will yield a contribution to the imaginary
 part of the classical action. The contribution of $P$
 to the tunneling rate is cancelled out when dividing the outgoing
 probability by the ingoing probability
 because $P$ is completely the same for both the outgoing and
 ingoing solutions. It is no need to solve
 the equations about the complex function $P$. So it is essential to
 work out the imaginary part of $R(\tilde{r})$.
 From the first and
 third formulae of the above equation, there will
 be a non-trivial solution for $A$ and $B$ if and only if the determinant
 of the coefficient matrix vanishes, which results
 \begin{eqnarray}
 \partial_{\tilde{r}}I=
 \frac{\omega}{(\tilde{r}^2/\tilde{r}_A^2-1)\sqrt{1-k\tilde{r}^2/a^2}}
 \left(H\tilde{r}\pm\sqrt{1-k\tilde{r}^2/a^2}\right)\;\;,
 \end{eqnarray}
 where the $+/-$ sign corresponds to the incoming/outgoing solutions respectively.
 It should be noted that the imaginary part of $R_\pm$ can be
 calculated using the above equation.
 Integrating the pole at the apparent horizon as in Refs.\cite{kerner1,mitra}, we have
 \begin{eqnarray}
 \textrm{Im}R_+&=&\textrm{Im}\int\frac{\omega}{(\tilde{r}^2/\tilde{r}_A^2-1)\sqrt{1-k\tilde{r}^2/a^2}}
 \left(H\tilde{r}+\sqrt{1-k\tilde{r}^2/a^2}\right)d\tilde{r}\nonumber\\
 &=&\pi\omega\tilde{r}_A\;\;,\nonumber\\
 \textrm{Im}R_-&=&\textrm{Im}\int\frac{\omega}{(\tilde{r}^2/\tilde{r}_A^2-1)\sqrt{1-k\tilde{r}^2/a^2}}
 \left(H\tilde{r}-\sqrt{1-k\tilde{r}^2/a^2}\right)d\tilde{r}\nonumber\\
 &=&0\;\;.
 \end{eqnarray}
 In the WKB approximation, the tunnelling probability
 is related to the imaginary part of the action as
 \begin{eqnarray}
 \Gamma=\frac{P_{\textrm{in}}}{P_{\textrm{out}}}=
 \frac{\textrm{exp}[-2(\textrm{Im}R_+ +\textrm{Im}P)]}
 {\textrm{exp}[-2(\textrm{Im}R_- +\textrm{Im}P)]}
 =\textrm{exp}[-2\pi\omega\tilde{r}_A]\;\;.
 \end{eqnarray}
 Comparing the tunneling probability and the thermal spectrum
 $\Gamma=\textrm{exp}[-\omega/T]$, Hawking temperature
 associated with the apparent horizon can be determined as
 \begin{eqnarray}
 T=\frac{1}{2\pi\tilde{r}_A}\;\;.
 \end{eqnarray}

 The Hawking temperature associated with the apparent horizon
 of FRW universe obtained by using the fermions tunneling method
 takes the same form as that obtained in \cite{cai1}.
 It should be noted that the Kodama observer is inside the apparent
 horizon. The results indicate that the Kodama observer does see a
 thermal spectrum with temperature $T=1/2\pi\tilde{r}_A$,
 which is caused by fermions tunneling from
 the outside apparent horizon to the inside apparent horizon.
 The results in the present paper together with that in \cite{cai1}
 fill in the gap existing in the literatures investigating the
 relationship between the first law of thermodynamics and Friedmann
 equations of FRW universe.

 On the other hand, the higher terms about $\omega$
 are neglected in our derivation and the expression (16) for tunneling
 probability implies the pure thermal radiation.
 When energy conservation is taken into account, the emission
 spectrum is no longer purely thermal and contains the higher terms about
 $\omega$.

\section*{ACKNOWLEDGEMENT}

 This work was supported by the National Natural Science Foundation
 of China and Cuiying Project of Lanzhou University.

 \end{document}